\documentclass[aps,prb,superscriptaddress,reprint,amsmath,amssymb]{revtex4-1}
\usepackage{graphicx}

\usepackage{psfrag, color}

\begin{document}

% Bauer, Flok, Muraki, PAn, ZnO, Nayak, Yuli, Rezayi, Simon
% check kappa, meff
% Barry, Gold, Shayegan, Wegschider, Pinczuk, Peterson, Jain, Papic

\title{Observation of an anomalous density-dependent energy gap of the \\
$\nu=5/2$ fractional quantum Hall  state in the low density regime}

\author{N. Samkharadze}
\email[]{Present address: QuTech and Kavli Institute of
NanoScience, Delft University of Technology, PO Box 5046, 2600 GA Delft, the Netherlands}
\affiliation{Department of Physics and Astronomy, Purdue University, West Lafayette, IN 47907, USA}
\author{Dohyung Ro}
\affiliation{Department of Physics and Astronomy, Purdue University, West Lafayette, IN 47907, USA}
\author{L.N. Pfeiffer}
\affiliation{Department of Electrical Engineering, Princeton University, Princeton, NJ 08544, USA}
\author{K.W. West}
\affiliation{Department of Electrical Engineering, Princeton University, Princeton, NJ 08544, USA}
\author{ G.A. Cs\'{a}thy}
\affiliation{Department of Physics and Astronomy, Purdue University, West Lafayette, IN 47907, USA}
\affiliation{Birck Nanotechnology Center, Purdue University, West Lafayette, IN 47907, USA}

\date{\today}

\begin{abstract}

We have studied the $\nu=5/2$ fractional quantum Hall state in a density-tunable sample
at extremely low electron densities. For the densities accessed in our experiment,
the Landau level mixing parameter $\kappa$ spans the  $2.52<\kappa<2.82$ range. In the vicinity of 
$5.8 \times 10^{10}$~cm$^{-2}$ or $\kappa = 2.6$ 
an anomalously large change in the density dependence of the energy gap is observed. 
Possible origins of such an anomaly are discussed, including a topological phase transition
in the $\nu=5/2$ fractional quantum Hall state.

\end{abstract}

\maketitle

\section{Introduction}

Unraveling the nature of the fractional quantum Hall state (FQHS) forming at the Landau level filling factor $\nu=5/2$
continues to be one of the challenging problems of many-body condensed matter physics. 
This FQHS was discovered \cite{willett} and studied in depth
in high quality two-dimensional electron gases (2DEGs) confined to GaAs/AlGaAs heterostructures
\cite{eisenstein88,pan99,panSSC01,miller07,dean08,choi08,radu08,dolev08,willett09,nuebler10,xia10,
zhang10,willett10,pan11,liu11,gamma,vivek11,kang11,dolev11,rhone11,lin12,nuebler12,muraki12,Mstern12,pinczuk13,
gamez13,willett13,jim13,pan14,baer,reichl14,alloy14,watson15,pinczuk15,pan15,gervais15,lin16,nodarkate}.
Other even denominator FQHSs in the GaAs/AlGaAs system develop at
$\nu=7/2$ \cite{eisen02,dean08,gamma,liu11b} and $\nu=2+3/8$ \cite{xia04,pan08,csa10}
and possibly related FQHSs have been observed at even denominators
in ZnO/MgZnO \cite{zno} and in bilayer graphene \cite{gr1,gr2,gr3,gr4}.
Since its discovery it was clear that, owing 
to the even denominator of the filling factor, the $\nu=5/2$ FQHS is not part of the sequence 
prescribed by the free composite
fermion theory \cite{jainCF,halp93}. Today this state is thought to belong to the Pfaffian universality class \cite{moore91}. 

The FQHS at $\nu=5/2$, however, admits theoretical descriptions which are topologically distinct from the Pfaffian.
Alternative candidate ground states for a FQHS at this filling factor are the anti-Pfaffian \cite{levin07,lee07},
the (3,3,1) Abelian state \cite{halperin83}, 
a variational wavefunction based on an antisymmetrized bilayer state \cite{park},
the particle-hole symmetric Pfaffian \cite{son, feldman},
a stripe-like alternation of the Pfaffian and anti-Pfaffian \cite{kun}, 
and other exotic states \cite{wen91,wen90}. An ongoing intense experimental effort is
not yet able to unambiguously discriminate between these gapped candidate states
\cite{radu08,kang11,dolev11,lin12,willett13,baer,lin16}.

The topologically distinct FQHSs at $\nu=5/2$ may compete in certain regions of the phase space.
For example, it was argued that
a direct topological phase transition between the Pfaffian and the anti-Pfaffian may occur \cite{levin07,lee07}.
Such a phase transition may be induced by tuning a parameter of the 2DEG.
One such tuning factor is the Landau level mixing parameter $\kappa$, a parameter typically
tuned by the electron density $n$ \cite{yoshi84}. 
It is known that by generating effective three-body terms in the interaction potential,
Landau level mixing (LLM) profoundly affects the Pfaffian and anti-Pfaffian states 
\cite{levin07,lee07,rezayi90,morf03,wojs06,wan08,peterson08,wang09,bishara09,wojs10,rezayi11,papic12,reza,troyer,yuli}.
Numerical work finds that these two states may compete
and thus a topological phase transition between the Pfaffian and the anti-Pfaffian is possible
\cite{rezayi90,morf03,wojs06,wan08,peterson08,wang09,bishara09,wojs10,rezayi11,papic12,reza,troyer,yuli}.
However, due difficulties stemming from the non-perturbative nature of the calculations and due
to limited computational resources, the stability region of the two phases and
details of a possible transition between the Pfaffian and the anti-Pfaffian remain obscure.
Nonetheless, the regime of large LLM occurring at low densities
has emerged as a region of interest for a possible phase transition in the $\nu=5/2$ FQHS.

The effect of a changing LLM of the $\nu=5/2$ FQHS has been studied only in a few experiments. 
%as $\kappa=E_c/\hbar \omega_C$ \cite{yoshi84}. Here $E_c=e^2/4\pi\epsilon l_B$ is the Coulomb energy, 
%$\hbar \omega_C$ the cyclotron energy, and $l_B=\sqrt{\hbar/eB}$ the magnetic length. At a given filling factor one
%can show that $\kappa \propto m^*/\sqrt{n}$, where $n$ is the electron density and $m^*$ the electronic effective mass.
At moderate LLM, i.e. with the LLM parameter $\kappa$ in the vicinity of $1$, there are no 
reports of any non-trivial phase transitions at $\nu=5/2$  \cite{nuebler10,pan11,liu11}.  
At $\kappa \gtrsim 2$ two interesting transitions at $\nu=5/2$ have been reported recently. The
closing and re-opening of the energy gap with density was interpreted as a 
spin transition in the $\nu=5/2$ FQHS \cite{pan14}.
In another experiment, pressure has drastically altered the ground state at $\nu=5/2$
 near $\kappa \simeq 2$ from a FQHS to a stripe phase \cite{nodarkate}.
Since in these experiments an in-situ tuning of the density was not possible,
subtle changes in sample properties are virtually impossible to detect.

Motivated by these ideas, we undertook a study of the energy gap of the $\nu=5/2$ FQHS in 
 the regime of very strong LLM with $\kappa > 2.5$. In order to track sample properties, 
we use a sample in which the density is tunable in-situ via a controlled illumination technique.
We report the observation of an anomalously sharp change in the 
density dependence of the energy gap of the $\nu=5/2$ FQHS
in the vicinity of $\kappa = 2.6$. 
We explored several possible interpretations of the observed anomaly and found
that this anomaly may be a first evidence
of a topological phase transition in the $\nu=5/2$ FQHS
between two distinct gapped ground states. 
%In contrast to the transition reported in Ref.\cite{pan14}, 
%in our experiment the energy gap does not close, but remains finite at the transition.

\section{Sample details and the cold illumination technique}

We studied a 65~nm wide symmetrically doped GaAs/AlGaAs quantum well sample 
with the density of $n=6.13 \times 10^{10}$~cm$^{-2}$ and the corresponding mobility 
of $\mu=9.1\times 10^6$~cm$^2$/Vs.
Heavily doped samples, such as ours, are often not responsive to 
the established technique of tuning the density by electrostatic gating. 
Instead of gating we use a low intensity illumination technique 
on an ungated sample. The preparation of sample state
starts with an illumination with a red light-emitting diode (LED) at 10~K using 1~mA excitation. 
This sets the density of our sample to it highest value $n=6.13 \times 10^{10}$~cm$^{-2}$.
To decrease the density, we apply a low excitation of the order of $1$~$\mu$A to the same LED for about 5 minutes while 
keeping the sample temperature close to 10~mK. As shown in Fig.1, the density reduction obtained 
as a result of successive low temperature illuminations scales with the product of the
LED excitation current and the time of illumination, a product we call the
integrated light intensity (ILI). As the density is reduced, the carrier mobility decreases
from its peak value of $\mu=9.1\times 10^6$~cm$^2$/Vs. This is shown in the inset of Fig.1. 
The illumination of samples is a known technique for state preparation  \cite{folk}
and it has been demonstrated that in certain cases it reduces the electron density \cite{tsuiLED}.

\begin{figure}[t]
% \begin{center}
 \includegraphics[width=0.85\columnwidth]{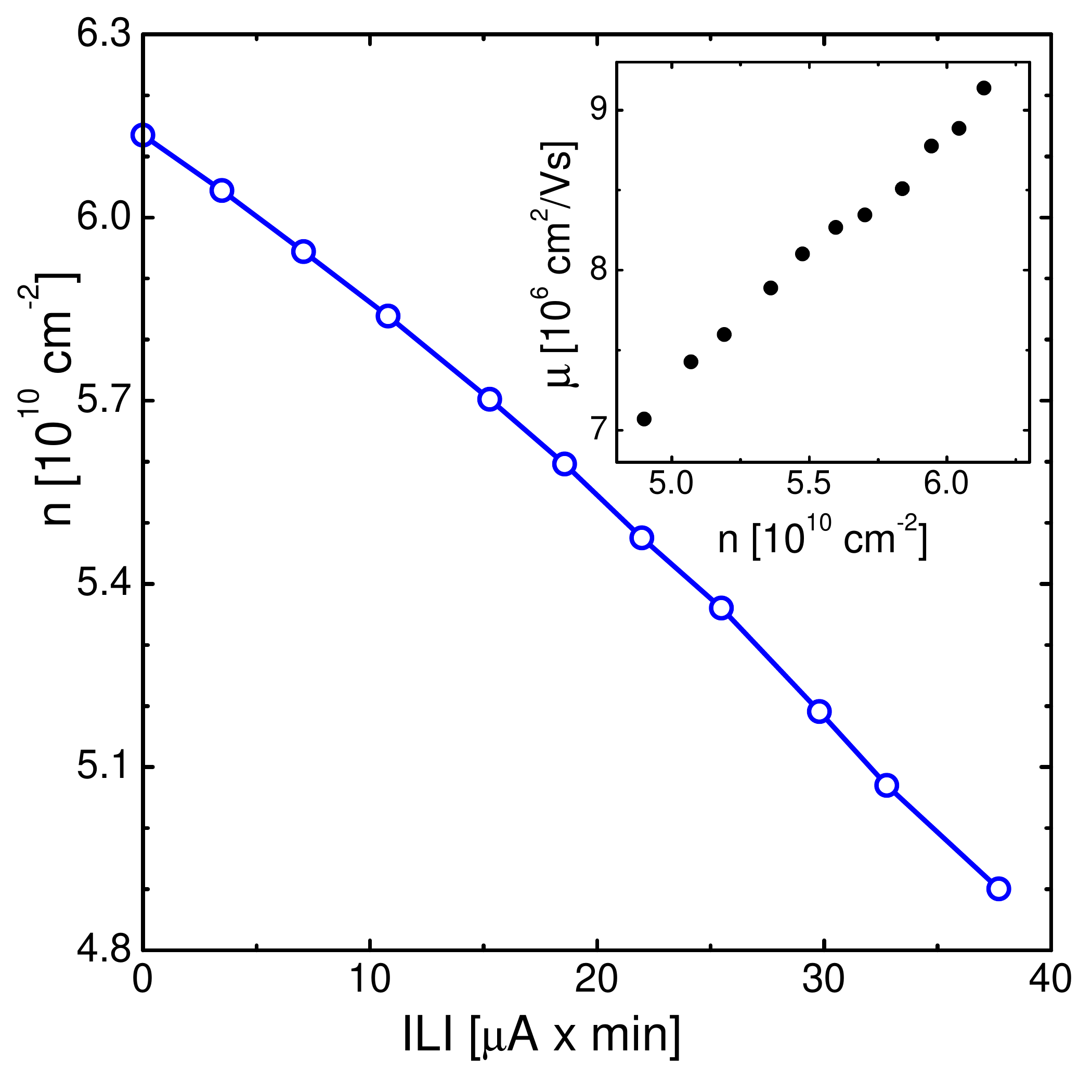}
% \end{center}
 \caption{
The electron density $n$ as a function of the integrated light intensity (ILI). 
The inset shows the mobility $\mu$ of the 2DEG as a function of the density.
}
\end{figure}

The LLM parameter $\kappa=E_c/\hbar \omega$ is the adimensional ratio of the Coulomb $E_c$ 
and the cyclotron $\hbar \omega$ energies  \cite{yoshi84}. Here $\omega = e B /m^*$ 
is the cyclotron frequency, $E_c = e^2 /(4 \pi \epsilon l_B)$, and  $l_B=\sqrt{\hbar/eB}$ is the magnetic length.
Since the density range of our sample is narrow, we approximate the 
effective mass $m^*$  with a constant $0.0670$ \cite{tan}.
Under this approximation $\kappa$ at any given filling factor scales with $1/\sqrt{n}$ \cite{yoshi84}.
A reduction in the electron density results therefore in an increase of $\kappa$.
Our experiment is performed in a density range for which $2.52<\kappa<2.82$. 

\section{Magnetoresistance data}

In Fig.2a we show the longitudinal magnetoresistance $R_{xx}$ in the vicinity of $\nu=5/2$ at three selected temperatures $T$ 
and at the highest achieved electron density $n=6.13 \times 10^{10}$~cm$^{-2}$. 
This is among the lowest densities to date at which the energy gap of the $\nu=5/2$ FQHS has been studied \cite{gamma,pan12PRL,pan14}.
The magnetic field $B$ in Fig.2a is such that the Landau level filling factor $\nu$ spans the 
$2<\nu<3$ range, commonly referred to as the lower spin branch of the second Landau level.
Here the filling factor is obtained from $\nu=h n/e B$, where $h$ is Planck's constant and $e$ is the elementary charge. 
At $\nu=5/2$ we observe a deep magnetoresistance minimum of less than 15~$\Omega$. 
Besides the FQHS at $\nu=5/2$, we also observe
strong FQHSs at $\nu=7/3$ and $11/5$, and there are indications of developing FQHSs at $\nu=8/3$, and $14/5$. 
Furthermore, in Fig.2a we notice three peaks in $R_{xx}$ at $B=1.10$, $1.04$, and $0.98$~T.
We associate these peaks with precursors at temperatures above the onset temperature
of the reentrant integer quantum Hall states $R2a$, $R2b$, and $R2c$, respectively \cite{deng}.

\begin{figure}[t]
% \begin{center}
 \includegraphics[width=1\columnwidth]{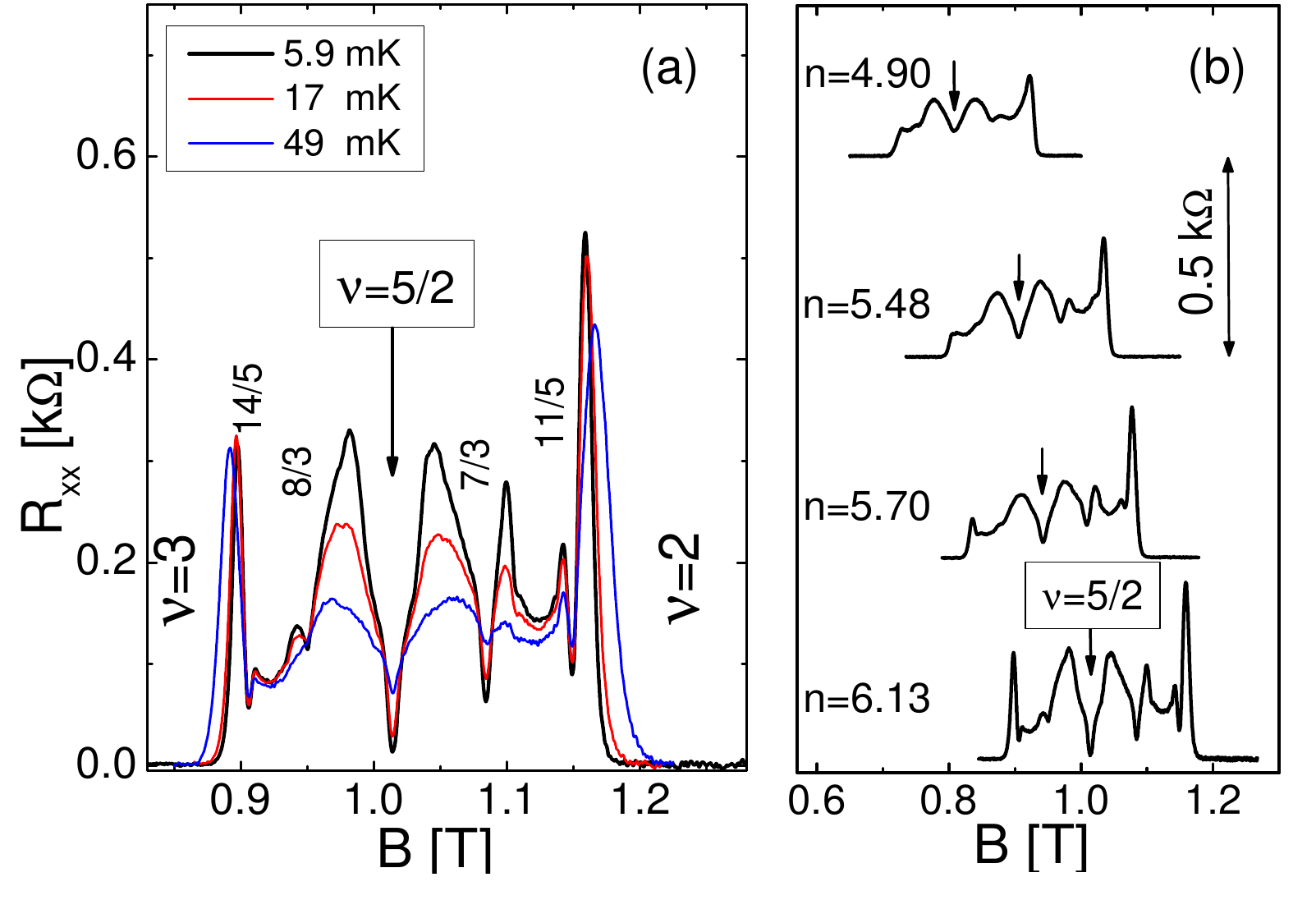}
% \end{center}
 \caption{
 (a) The magnetoresistance at three different temperatures at the
 largest achieved sample density n=$6.13 \times 10^{10}$~cm$^{-2}$. The integer
 and fractional quantum Hall states are marked by their filling factor $\nu$.
 (b) The magnetoresistance at $T=5.9$~mK at several different 
 electron densities. The labels are densities in units of $10^{10}$~cm$^{-2}$. 
 Arrows mark the position of the $\nu=5/2$ FQHS.
}
\end{figure}

The activation energy gap $\Delta_{5/2}$ of the $\nu=5/2$ FQHS
is extracted from the temperature dependence of the magnetoresistance using the 
$R_{xx} \propto \exp(-\Delta_{5/2}/2T)$ relation. Since in experiments one typically measures
the bath temperature, accurate gap measurements require a good thermalization 
of the electrons to the bath. To achieve a good thermalization, we
use a He-3 immersion cell \cite{cell}.
The bath temperature is monitored using a quartz tuning fork thermometer. Such a thermometer is
well suited for low temperature measurements in strong magnetic fields since it is immune to 
rf heating and it is independent of the $B$-field \cite{cell}.
The Arrhenius plots of the magnetoresistance at $\nu=5/2$ for several different electron densities are shown in Fig.3.
The energy gap is then extracted from the slope of the linear part of the data. We note that as
the $\nu=5/2$ FQHS becomes more fragile with the reduction of the density, the range of the linear fit
is reduced. However, similarly to other work on states with small energy gaps
\cite{nuebler10,pan14}, we assume that the energy gap can be extracted from the linear part of the data.

\begin{figure}[t]
% \begin{center}
 \includegraphics[width=1\columnwidth]{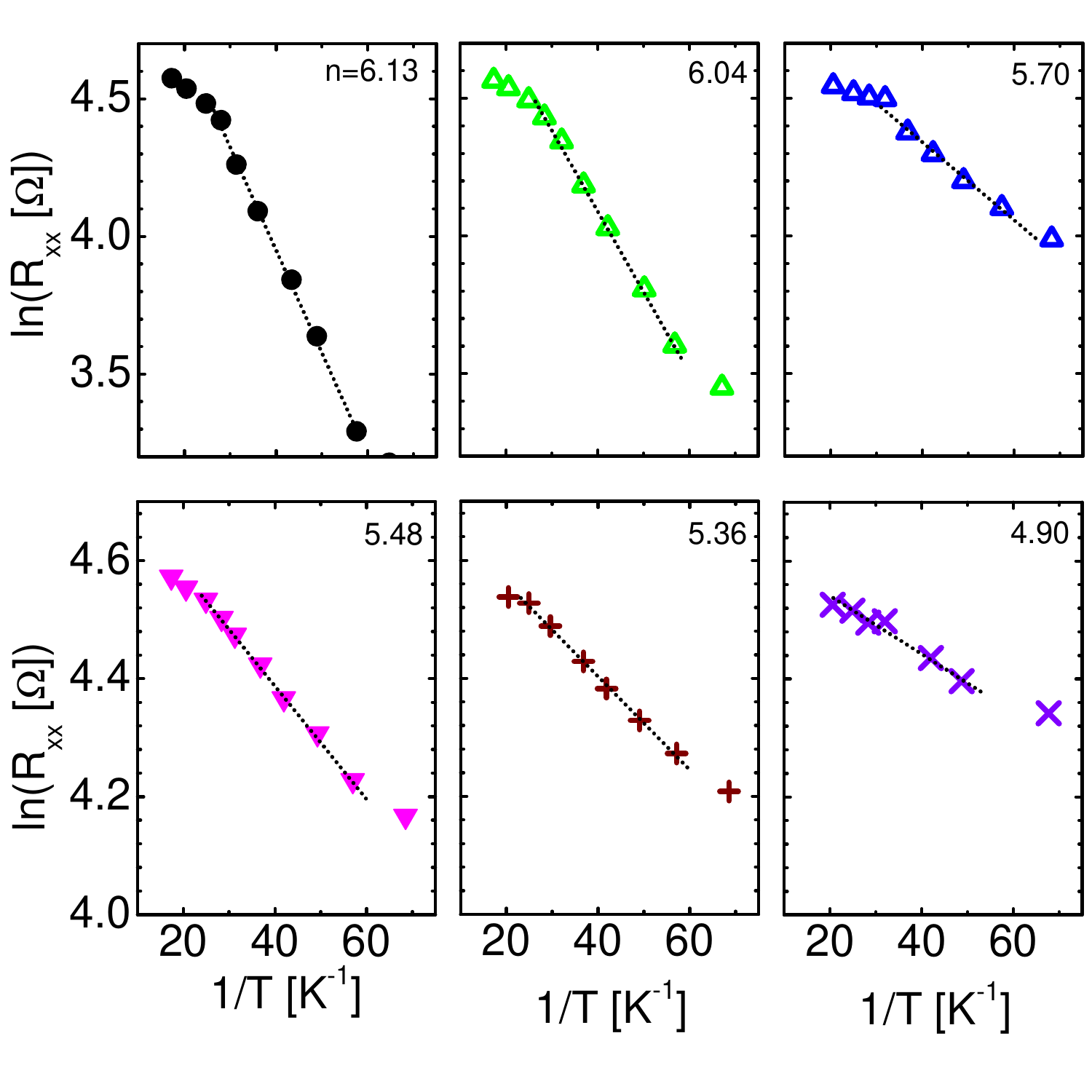}
% \end{center}
 \caption{
 Arrhenius plots for $R_{xx}$ at $\nu=5/2$ filling factor at several representative densities.  
 The dotted lines are fits used to extract the energy gap. Each panel shows the electron density
 in units of $10^{10}$cm$^{-2}$.
}
\end{figure}

At the highest prepared density $n=6.13 \times 10^{10}$~cm$^{-2}$, the gap of $\nu=5/2$ FQHS is $\Delta_{5/2}$=80~mK. 
A similar measurement of the $\nu =7/3$ FQHS (not shown) results in $\Delta_{7/3}$=27~mK. 
The energy gap at this density of the $\nu=7/3$ FQHS is, therefore,
a factor of 3 smaller than that of the $\nu=5/2$ FQHS. This result is surprising since 
in numerous samples the energy gaps at $\nu=7/3$ and $\nu=5/2$ are either comparable  
\cite{pan99,xia10, pan08,csa10,gamma,dean08,nuebler10,choi08,miller07,liu11}
or the gap at $\nu=7/3$ is larger than that at $\nu=5/2$ \cite{liu11}.
For example, the gaps of these two states  
in a sample of low density $n=8.3 \times 10^{10}$~cm$^{-2}$ are $\Delta_{5/2}=88$~mK and $\Delta_{7/3}=81$~mK  \cite{gamma}.
We ascribe such a reduction of the energy gap of the $\nu=7/3$ FQHS as compared to that of the
$\nu=5/2$ FQHS to the large LLM present in our sample.

\section{Density dependence of the energy gap at $\nu=5/2$}

As the density is decreased using the cold illumination technique, both the $\nu=5/2$ and $7/3$ FQHS become
more fragile. This is seen in Fig.2b where we show the evolution of $R_{xx}$ with the magnetic field
as measured at $T=5.9$~mK at different densities. Consistent with this behavior, $\Delta_{5/2}$
exhibits a decreasing trend with a decreasing density. This trend is captured in Fig.4. 
Such a suppression of $\Delta_{5/2}$ with a decreasing density
is expected \cite{morf03,wojs06,wojs10} and it has experimentally been 
observed in the regime in which only the lowest electrical subband is populated
\cite{panSSC01,gamma,nuebler10,liu11,pan11,pan14}.
A contrasting behavior of precipitous collapse of $\Delta_{5/2}$ with an increasing density \cite{liu11}
has only been observed with the polulation of the second electrical subband \cite{xia10,liu11,papic12}.  
The absence of any beating in the Shubnikov-de Haas oscillations and a lack of reduction of
the mobility with an increasing density show that in our sample we do not populate the second electric subband
\cite{panSSC01, stormer82}. We note that a  linear extrapolation of our lowest density data from Fig.4 shows that
$\Delta_{5/2}$ vanishes near the extrapolated value of $n \simeq 4.5 \times 10^{10}$~cm$^{-2}$, a 
value which is in a reasonable agreement with other data measured at low densities \cite{gamma,pan14}. 

\begin{figure}[t]
 \includegraphics[width=0.9\columnwidth]{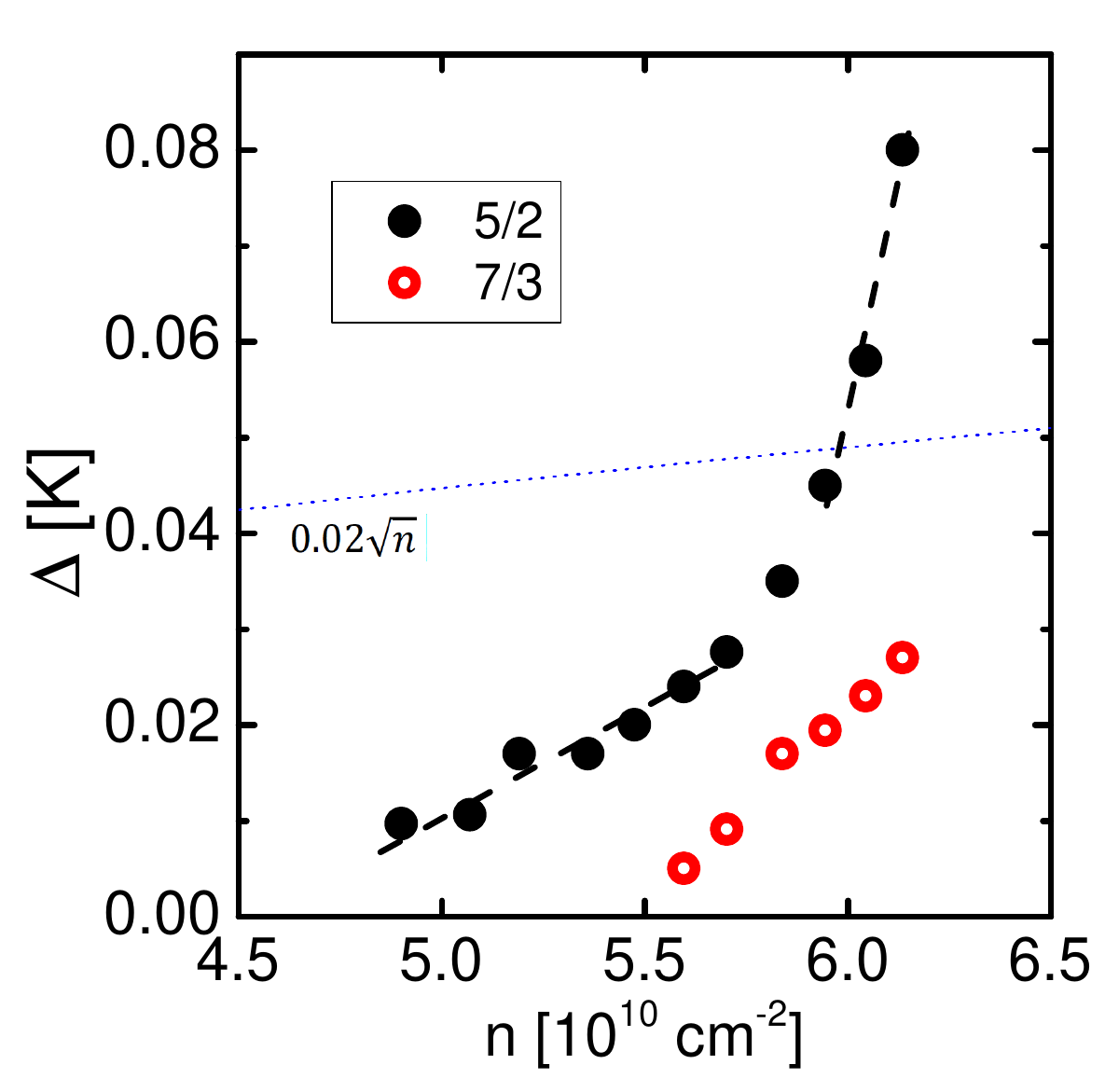}
 \caption{
 Dependence of the energy gaps of the $\nu=5/2$ and $\nu=7/3$ FQHSs on the density. 
 The top scale shows the LLM parameter $\kappa$ calculated at $\nu=5/2$.
 Dashed lines are linear fits to the data. Near $\kappa \approx 2.6$ we observe 
 an anomaly indicated by a very large change in the slope. The dotted line illustrates
 a $\sqrt{n}$ functional dependence.
}
\end{figure}

While the decreasing trend of $\Delta$ with a decreasing $n$ was expected, the funtional
dependence of $\Delta$ versus $n$ exhibited in Fig.4 is highly unusual.
In a disorder-free environment the energy gap at a given filling factor is expected 
to scale with the Coulomb energy $E_C$ and, therefore with $\sqrt{n}$. 
As seen in Fig.4, the measured dependence of $\Delta$ on $n$ is clearly not captured by such a 
simple $\sqrt{n}$ functional dependence. However, in addition to the Coulomb term
proportional to $\sqrt{n}$ one also has to consider the disorder broadening of the energy levels
\cite{morf03}, a parameter with an unknown density dependence 
\cite{panSSC01,gamma,nuebler10,liu11,pan11,pan14}.

The most salient feature of our data is the change in the slope of the $\Delta_{5/2}$ versus $n$ curve
which is apparent in Fig.4 in a positive curvature close to the density 
of $5.8 \times 10^{10}$~cm$^{-2}$ or $\kappa = 2.6$.
At the highest densities $\Delta_{5/2}$ decreases very steeply with a decreasing $n$. In contrast,
at the lowest densities $\Delta_{5/2}$ has a more gentle slope. 
By fitting the low and high ends of our $\Delta_{5/2}$ versus $n$ plot with straight lines, we find that 
in Fig.4 the slope $\partial \Delta_{5/2} / \partial n$ changes by a factor 10, which is anomalously large.
Remarkably, the anomalous change in the slope of the $\Delta_{5/2}$ versus $n$
curve shown in Fig.4 takes place as the density is changed by less than 5\%. 
This abrupt change in the slope over a narrow density range is suggestive of a phase transition.

\section{Discussion}

A positive curvature of the $\Delta_{5/2}$ versus $n$ curve shown in Fig.4 associated with the 
large change in the slope is not present in data sets from Refs.\cite{panSSC01,nuebler10,pan11,pan14}. 
A positive curvature of the $\Delta_{5/2}$ versus $n$ dependence is, however, seen 
in the two data sets at relatively high densities from Ref.\cite{liu11}. Nonetheless, we think that 
the positive curvature observed in our experiment and that in Ref.\cite{liu11} must have a different origin.
Indeed, when comparing our data to those from Ref.\cite{liu11}, there are two important differences. 
First, the two experiments are performed at the opposite ends of experimentally accessible 
LLM for electron samples. Thus, in contrast to Ref.\cite{liu11}, in our sample the Coulomb energy is 
the dominant energy scale. 
Second, the relationship between the 5/2 and 7/3 FQHSs in our sample is very different 
from that from Ref.\cite{liu11}.  In Ref.\cite{liu11} $\Delta_{7/3}$ measured for the 30~nm 
quantum well sample, is either similar to or greatly exceeds $\Delta_{5/2}$, whereas in our sample
$\Delta_{7/3}$ is considerably lower than $\Delta_{5/2}$, by at least a factor 3.

One possible interpretation of our data is a transition from fully to partially polarized state at $\nu=5/2$ as 
the density is lowered \cite{eisenstein89,pan12prb,pan14,pan12PRL}.  
A recent experiment probing the $\nu=5/2$ found that as the density is lowered, the energy 
gap at $\nu=5/2$ decreases, it nearly closes, then it increases at the lowest densities, 
a behavior which was interpreted as evidence of a spin transition \cite{pan14}.
Our data shown in Fig.4 is quite different from that in Ref.\cite{pan14} since in contrast to the latter,
in our experiment $\Delta_{5/2}$ does not close at the transition point. 
We thus think that our data cannot be interpreted as a spin transition in the $\nu=5/2$ FQHS. 

The observed positive curvature shown in Fig.4
in the density dependence of $\Delta_{5/2}$  may be caused by the effects of the disorder.
It is known that disorder broadening of the energy levels plays an important role in the measured
energy gaps of the FQHSs \cite{morf03} and  the positive curvature shown in Fig.4 may be due
to an anomalous dependence of the disorder broadening on the density. 
% e wonder if our illumination procedure causes the anomalous behavior.
In the past, following the analysis proposed in Ref.\cite{morf03}, 
we extracted the intrinsic gap and the disorder broadening for the $\nu=5/2$ FQHS in several samples \cite{gamma}.
In the absence of a measurable gap at $\nu=7/2$, the same type of analysis cannot be applied to the present sample
in which density is tuned by the illumination technique described earlier. 
The density dependence of the disorder broadening remains thus experimentally inaccessible.
Nonetheless we think that had the disorder been the cause of the anomaly observed at $\nu=5/2$, 
a similar effect would also be present at $\nu=7/3$. However, as already discussed, 
the density dependence of the gaps at $\nu=5/2$ and $\nu=7/3$ is very different. Furthermore, the smooth
density dependence of the mobility shown in the inset of Fig.1 strengthens the argument that disorder 
effects are unlikely to drive the anomaly seen in Fig.4 at $\nu=5/2$. Nonetheless, for a better understanding,
further studies of the influence of disorder generated with illumination will be neccessary.

A LLM induced phase transition between two distinct gapped ground states at $\nu=5/2$ is another interpretation of our data.
As discussed in the second paragraph of this Article, there are several topologically distinct candidate ground states for the
$\nu=5/2$ FQHS: the Pfaffian \cite{moore91}, anti-Pfaffian \cite{levin07,lee07}, the (3,3,1) Abelian state \cite{halperin83}, 
a variational wavefunction based on an antisymmetrized bilayer state \cite{park}, the particle-hole symmetric 
Pfaffian \cite{son, feldman}, a stripe-like alternation of the Pfaffian and anti-Pfaffian \cite{kun}, 
and other exotic states \cite{wen91,wen90}. One thus may consider a potentially large number of topological 
phase transitions between the pairwise distinct candidate ground states. We note that
it is generally believed that at a topological phase transition in a disorder-free sample 
the energy gap has to vanish. However, one could envision situations in which
the gap does not fully close at a topological phase transition.
For example, in a realistic sample some rounding of the transition may occur 
and the gap therefore may not fully close. Furthermore, there may be topological phase transitions
for which the gap of neutral exitations closes, but the gap of the charge excitations does not.
%For such a transition the measured energy gap not detectable by charge measurement as a vanishing energy gap. 

LLM is known to be the leading generator of three-body terms $V^{(3)}$ in the effective interaction
and may therefore infuence which ground state is stabilized.
%the sign of this three-body potential determines whether the preferred ground state is the Pfaffian or the anti-Pfaffian. 
However, due to difficulties arising from a large Hilbert space,
the inclusion of LLM in numerical calculations remains a formidable task
\cite{rezayi90,morf03,wojs06,wan08,peterson08,wang09,bishara09,wojs10,rezayi11,papic12,reza,troyer,yuli}. 
Recent numerical work reports that at $\nu=5/2$ the anti-Pfaffian is stabilized \cite{rezayi11,papic12}
at moderate LLM $\kappa \approx 1$, there are indications
that the stability of the Pfaffian may be enhanced \cite{wojs10} near $\kappa=2$, and it is thought that
a transition between the Pfaffian and the anti-Pfaffian is possible as LLM is tuned \cite{levin07,lee07,wang09,bishara09}.
However, the value of $\kappa_{crit}$ at such a transition is unknown.   %  \cite{bishara09, wang09}. 
We may obtain an estimate from a recent calculation  \cite{bishara09} of the 
$m=3$ term of the three-body potential $V^{(3)}_{3,3/2} \approx -0.0147\kappa+0.006\kappa^2$ which
vanishes at $\kappa_{crit} \approx 2.5$.
The abrupt change in the slope of $\Delta_{5/2}$ versus $n$ curve shown in Fig.4 in the vicinity 
of $\kappa = 2.6$ is very close to the above estimate. 
We note, however, that in lack of knowledge of the contributions from  particle-hole symmetry breaking terms 
\cite{bishara09, wojs10} other than $V^{(3)}_{3,3/2}$, the estimate of $\kappa_{crit}$ remains quite crude.  
In addition to the the Pfaffian to anti-Pfaffian transition, topological phase transitions may also
be allowed between other candidate ground states of the $\nu=5/2$ FQHS 
\cite{halperin83,park,son,feldman,kun,wen91,wen90}. One example is
the transition from the Pfaffian to the (3,3,1) Abelian state studied in a numerical experiment \cite{biddle}.
We thus think that a topological phase transition, such as the transition between the Pfaffian and the
anti-Pfaffian expected at large $\kappa$\cite{levin07,lee07,wang09,bishara09}, the
transition from the Pfaffian to the (3,3,1) Abelian state \cite{biddle}, or a topological phase transition 
of a different kind \cite{tapash},
remains an exciting possible interpretation of our data. 
The recently observed rich phase diagram exhibiting several even denominator FQHS in bilayer graphene \cite{gr3,gr4}
opens up the prospects of observing related topological phase transitions in systems other than GaAs.

%While the existence of a topological phase transition is most often associated with the closing of the energy gap, 

Numerical work \cite{papic12} suggests that a positive curvature in the $\Delta_{5/2}$ versus $n$ curve 
may be caused by a novel type of mixing of the Landau levels associated with different electric subbands \cite{xia10,liu11}.
Although this new type of mixing may play an important role in the physics of the $\nu=5/2$ FQHS,
we think that the results of the calculation \cite{papic12} do not strictly apply to our sample.
Indeed, mixing with the excited Landau levels of the lowest electric subband are not included
in Ref.\cite{papic12} but, as we argued earlier, it plays an important role in our experiment.

\section{Conclusions}

In conclusion, we have studied the density dependence of the fractional quantum Hall state at 
$\nu=5/2$ in the regime of extremely low densities and, hence, large LLM. We have observed
an anomaly in the density dependence of the energy gap of this state. The observed anomaly 
is consistent with expectations of a topological 
phase transition between two topologically distinct FQHSs at $\nu=5/2$. 
We also analyzed other possible origins of the observed anomaly in the $\nu=5/2$ FQHS but 
found that a spin transition, effects of the second electric subband in the confining potential, 
and effects of the disorder are unlikely to account for our observations.

\section*{Acknowledgments}

We thank M. Peterson, Z. Papi\'c, and M. Shayegan for helpful discussions. The work at Purdue was 
supported by the NSF grants DMR-1207375 
and DMR-1505866. L.N.P. and K.W.W. acknowledge the Gordon and Betty Moore Foundation
Grant No. GBMF 4420, and the National Science Foundation MRSEC Grant No. DMR-1420541.

\end{document}